\documentclass[aps,twocolumn,superscriptaddress,amsmath,amsfonts,showkeys,floatfix]{revtex4}

\usepackage{epsfig,psfrag}
\usepackage{CJK}

\usepackage{graphicx}
\usepackage{dcolumn}
\usepackage{bm}
\usepackage{amssymb}
\usepackage{natbib}

\begin{document}
\title{Theory of all-coupling angulon for molecules rotating in many-body environment}
\author{Yi-Yan Liu}
\affiliation{Tianjin Key Laboratory of Low Dimensional Materials Physics and Preparing Technology, Department of Physics, School of Science, Tianjin University, Tianjin 300354 China}
\author{Yu Cui}
\affiliation{Tianjin Key Laboratory of Low Dimensional Materials Physics and Preparing Technology, Department of Physics, School of Science, Tianjin University, Tianjin 300354 China}
\author{Xiao-Zhe Zhang}
\affiliation{Tianjin Key Laboratory of Low Dimensional Materials Physics and Preparing Technology, Department of Physics, School of Science, Tianjin University, Tianjin 300354 China}
\author{Ran-Bo Yang}
\affiliation{Tianjin Key Laboratory of Low Dimensional Materials Physics and Preparing Technology, Department of Physics, School of Science, Tianjin University, Tianjin 300354 China}
\author{Zhi-Qing Li}
\affiliation{Tianjin Key Laboratory of Low Dimensional Materials Physics and Preparing Technology, Department of Physics, School of Science, Tianjin University, Tianjin 300354 China}
\author{Zi-Wu Wang}
\email{wangziwu@tju.edu.cn}
\affiliation{Tianjin Key Laboratory of Low Dimensional Materials Physics and Preparing Technology, Department of Physics, School of Science, Tianjin University, Tianjin 300354 China}
\begin{abstract}
The formation of angulon, stemming from the rotor (molecule or impurity) rotating in the quantum many-body field, adds a new member in the quasiparticle's family and has aroused intensively interests in multiple research fields. However, the analysis of the coupling strength between the rotor and its hosting environment remains a challenging task both in theory and experiment. Here, we develop the all-coupling theory of the angulon by introducing an unitary transformation, where the renormalization of the rotational constants for different molecules in the helium nanodroplets are reproduced, getting excellent agreement with the collected experimental data during the past decades. Moreover, the strength of molecule-helium coupling and the effective radius of the solvation shell corotating along with the molecular rotor could be estimated qualitatively. This model not only provides the significant enlightenment for analyzing the rotational spectroscopy of molecules in the phononic environment, but also provides a new method to study the transfer of the phonon angular momentum in angulon frame.
\end{abstract}
\keywords{angulon, superfluid helium, rotational constant, phonon angular momentum}
\maketitle

\section{Introduction}
A new member of the quasiparticle's family---$angulon$, representing the entity of the quantum rotor (the rotational molecule or impurity) dressed by the quantum many-body excitation, proposed by Schmidt and Lemeshko in 2015\cite{w1}, which has been arising more and more interests in many research areas during the past years\cite{w2,w3}, because it knocks the door to explore a series of novel phenomena, such as the angular self-localization effect\cite{wf1,wf2}, the possible realization of magnetic monopoles\cite{wf3} and the nontrivial topology of the rotational molecule driven by periodic far-off-resonant laser pulses\cite{wf4}.

In fact, different types of particles rotating in the many-body environment have been extensively studied in the past decades. For instance, impurities immersed in superfluid helium\cite{w8,w9,w10} and Bose-Einstein-condensate\cite{w11,w12}, clusters or organic molecules rotating in these cage-like structures including the hybrid metal halide perovskites\cite{w14}, fullerene\cite{wk1,wk2}, carbon nanotube\cite{wk3,wk4} and so on. Especially, with its stable ultracold isolated matrix, the superfluid helium ($^4{\rm{He}}$) nanodroplets provides an ideal environment to investigate the molecular fine spectroscopy and dynamics, hindering some external distraction\cite{w9,w15,w16,w17,w18,w19,w20,w21}. Though superfluid helium is like a refrigerator, the coupling between molecules and the helium bath is inevitable\cite{w9,w18}, resulting in the redistribution of angular momenta between them, which have been extensively proved by the asymmetric rotational line-shapes of infrared spectra\cite{w10,w15,w16,w17,w18,w19,w20,w21,w24,w25,w27,w28,w29,w30,w31}, the Stark-like splitting of spectra\cite{wr1,wr2} and the local solvation shell induced revival time of molecule\cite{w10,w33,w34,w35,w36,w37,wu1} in many experiments. In other words, the coupling between rotating molecules and helium can cause the rotational fine structures, which could be reflected directly by the renormalization of rotational constants due to the angular momenta transfer of many-body environment\cite{wk5}, enlightening by the concept of effective mass in polaron picture\cite{wt1,wt2}. Many experiments have also pointed out that the rotational constants change significantly respect to the free rotational motion for different molecular species in many-body systems\cite{w8,w18}, implying the coupling strength between them varies in the very large scale. But the analysis of this coupling strength in the whole range remains a complicated task both in theory and experiment. Within the frame of the angulon model, Lemeshko proposed a phenomenological method to study the renormalization of the effective rotational constants for molecules in the weak- and the strong-coupling limit, respectively\cite{wm1}. They found, however, that there are overestimation or underestimation for some molecules comparing with experimental measurements. Meanwhile, their method depends on the classification of coupling strength and the corresponding phenomenological parameters for the weak- and the strong-coupling limit, respectively, rendering it lacking of the general applicability. Therefore, just what they pointed out an all-coupling model for this problem is urgent.

In this paper, we develop an all-coupling angulon model to study different molecules rotating in the superfluid helium environment by introducing an unitary transformation, where a very simple and effective formula for the renormalization of the rotational constant is derived. With the help of this formula, we could reproduce the effective rotational constants for molecules from light- to intermediate- and heavy-mass species, getting the excellent agreement with the experimental data collected in the past several decades. Differing from some fixed values from potential energy surface\cite{wm1,wx1} between the molecules and helium for the coupling strength and two phenomenological parameters were adopted in the previous model, our simulation process depends on two intrinsic parameters: the strength of the molecule-helium coupling and the effective radius of the solvation shell corotating along with the molecules. One can find that the very broad range of the renormalization effect for the rotational motion of molecules would be reproduced successfully when these two parameters with the appropriate values, allowing this model could be expanded to study general systems that particles rotating in quantum many-body environment.

\section{Theoretical Model}
We begin with the angulon Hamiltonian describing the coupling of a rotating molecule with a phononic bath\cite{w1,wf1,wf2,wm1,wp1}:
\begin{equation}
\hat H = {{\hat H}_{\rm{k}}} + {{\hat H}_{{\rm{ph}}}} + {{\hat H}_{{\rm{int}}}},
\end{equation}
where, ${{\hat H}_{\rm{k}}}$ corresponds to the rotational kinetic energy of the molecule
\begin{equation}
{{\hat H}_{\rm{k}}}=B{\hat {\bf J}^2},
\end{equation}
with $\hat {\bf J}$ being the angular momentum operator, $B ={1 \mathord{\left/
 {\vphantom {1 {2I}}} \right. \kern-\nulldelimiterspace} {2I}}$ is the free rotational constant, where $I$ is the molecular moment of inertia. Without an external bath, the rotational eigenstates $\left| {LM} \right\rangle$ are labeled by the angular momentum $L$ and its projection $M$ onto the laboratory-frame $z$ axis, and energies ${E_L} = BL\left( {L + 1} \right)$ correspond to the $(2L+1)$-fold degenerate for these rotational states\cite{w34,wa1}.

The second Hamiltonian ${{\hat H}_{\rm{ph}}}$ represents the energy of the phononic bath, arising from the superfluid helium, and can be written as
\begin{equation}
{{\hat H}_{\rm{ph}}}=\sum\limits_{k\lambda \mu } {{\omega _k}} \hat b_{k\lambda \mu }^\dag {\hat b_{k\lambda \mu }},
\end{equation}
where the corresponding creation $\hat b_{\bf k}^\dag$ and annihilation ${{{\hat b}_{\bf k}}}$ operators are expressed in the spherical basis, $\hat b_{k\lambda \mu }^\dag$ and ${{{\hat b}_{k\lambda \mu }}}$\cite{w1,wf1,wf2,wp1}, respectively. Here, $\left| {\bf k} \right| = k$ is the wave vector of phonon mode, while $\lambda $ and $\mu $ define, respectively, the quantum number of the phonon angular momentum and its projection onto the laboratory $z$ axis. Here, the dispersion relation ${\omega _k} = \sqrt {{\epsilon _k}\left( {{\epsilon _k} + 2{g_{bb}}n} \right)} $ is adopted for the superfluid helium with ${\epsilon _k} = {k^2}/{2m}$, where $m$, $n$ and ${g_{bb}}$  denote the mass, density of helium and the parameter of helium-helium interaction, respectively\cite{w1,w2,wf1,wf2,wf3,w34,wm1,wp1}. Additionally, $\hbar  \equiv 1$ is set throughout this paper.

 The third term describing the couplings between the molecule and the phononic bath is given by
\begin{equation}
{{\hat H}_{{\rm{int}}}}= \sum\limits_{k\lambda \mu } {{U_\lambda }\left( k \right)} \left[ {Y_{\lambda \mu }^ * \left( {\hat \theta ,\hat \phi } \right)\hat b_{k\lambda \mu }^\dag  + {Y_{\lambda \mu }}\left( {\hat \theta ,\hat \phi } \right){{\hat b}_{k\lambda \mu }}} \right],
\end{equation}
where $Y_{\lambda \mu }\left( {\hat \theta ,\hat \phi } \right)$ are spherical harmonics and $\sum\nolimits_k { \equiv \int {dk} }$. The angular momentum-dependent coupling strength ${{U_\lambda }\left( k \right)}$ depends on the microscopic details of the two-body interaction between the molecule and the phonons. For a simple case that a linear rotor immersed into the Bose gas, the coupling matrix is approximated by\cite{w1,wm1,wp1}
\begin{equation}
{U_\lambda }\left( k \right) = {u_\lambda }{\left[ {\frac{{8{k^2}{\epsilon _k} n }}{{{\omega _k}\left( {2\lambda  + 1} \right)}}} \right]^{ {1 \mathord{\left/
 {\vphantom {1 2}} \right.
 \kern-\nulldelimiterspace} 2}}}\int {dr{r^2}{f_\lambda }\left( r \right){j_\lambda }\left( {kr} \right)},
\end{equation}
with ${u_\lambda }$ and ${f_\lambda }\left( r \right)$ represent the strength and shape of the coupling potential, respectively, in the $\lambda_{th}$ channel of the phonon angular-momentum. In general, the Gaussian-type shape factor ${f_\lambda }\left( r \right) = {\left( {2\pi } \right)^{ - {3/ 2}}}{e^{ -{{{r^2}}/ {2R_\lambda ^2}}}}$ is adopted\cite{w1,wm1,wp1}, where $R_\lambda$ denotes the effective radius of the solvation shell of the phononic bath, corotating with the rotating molecules, which reflects the range of a local density deformation of helium and plays a crucial role to describe the coupling between molecule and superfluid helium; $j_{\lambda}(kr)$ is the spherical Bessel function.

Inspired by the all-coupling polaron model\cite{wp2,wp3,wp4}, we introduce an unitary transformation
\begin{equation}
\hat S = exp\left[ {\sum\limits_{k\lambda \mu } {[F_{k\lambda \mu }^{\rm{*}}\left( {\hat \theta ,\hat \phi } \right){{\hat b}_{k\lambda \mu }} - {F_{k\lambda \mu }}\left( {\hat \theta ,\hat \phi } \right){\hat b_{k\lambda \mu }^\dag}]} } \right],
\end{equation}
where ${{F_{k\lambda \mu }}\left( {\hat \theta ,\hat \phi } \right)}$ is the variational function, satisfying the relation of
\begin{equation}
{F_{k\lambda \mu }}\left( {\hat \theta ,\hat \phi } \right) = {U_\lambda }\left( k \right)\mathbb{Z} Y_{\lambda \mu }^*\left( {\hat \theta  ,\hat \phi } \right) + {U_\lambda }\left( k \right)\mathbb{W},
\end{equation}
with $\mathbb{Z}$ and $\mathbb{W}$ being the variational parameters.

Performing the transformation $\tilde H = {S^{ - 1}}\hat HS$ for the angulon Hamiltonian, we can get
\begin{widetext}
\begin{eqnarray}
&&{S^{ - 1}}{\hat H_{\rm{k}}}S \nonumber\\
&=& B{\left( {{{\hat J}_0} - {{\hat M}_0}} \right)^2} + 2B{{\hat P}_0} \left( {{{\hat J}_0} - {{\hat M}_0}} \right) + B\left( {{{\hat J}_0} + {{\hat P}_0} - {{\hat M}_0}} \right)+B{{\hat P}_0}^2 - 2B\left( {{{\hat J}_{ - 1}} - {{\hat M}_{ - 1}}} \right)\left( {{{\hat J}_{ + 1}} - {{\hat M}_{ + 1}}} \right)\nonumber\\
&& - 2B{{\hat P}_{ - 1}} \left( {{{\hat J}_{ + 1}} - {{\hat M}_{ + 1}}} \right)  - 2B{{\hat P}_{ + 1}}  \left( {{{\hat J}_{ - 1}} - {{\hat M}_{ - 1}}} \right)+ B\sum\limits_{k\lambda \mu } {{U_\lambda }\left( k \right)} {\mu ^2}\left[ {{Y_{\lambda \mu }}\left( {\hat \theta ,\hat \phi } \right){{\hat b}_{k\lambda \mu }} + Y_{\lambda \mu }^ * \left( {\hat \theta ,\hat \phi } \right)\hat b_{k\lambda \mu }^\dag } \right]\mathbb{Z}\nonumber\\
&&-2B{{\hat P}_{ - 1}} {{\hat P}_{ + 1}}+ 2B\sum\limits_{k\lambda \mu } {{U_\lambda }\left( k \right)} \left[ {\frac{{\lambda \left( {\lambda  + 1} \right) - \mu \left( {\mu  + 1} \right)}}{2}{Y_{\lambda \mu }}\left( {\hat \theta ,\hat \phi } \right){{\hat b}_{k\lambda \mu }} - \frac{{\lambda \left( {\lambda  + 1} \right) - \mu \left( {\mu  - 1} \right)}}{2}Y_{\lambda \mu }^*\left( {\hat \theta ,\hat \phi } \right)\hat b_{k\lambda \mu }^\dag } \right]\mathbb{Z}, \nonumber\\
\end{eqnarray}
\end{widetext}
\begin{eqnarray}
&&{S^{ - 1}}{{\hat H}_{\rm bos}}S \nonumber\\
&=&\sum\limits_{k\lambda \mu } {{\omega _k}} {\rm{ }}\left[ \hat b_{k\lambda \mu }^\dag {{\hat b}_{k\lambda \mu }} - \hat b_{k\lambda \mu }^\dag {F_{k\lambda \mu }}\left( {\hat \theta ,\hat \phi } \right)\right. \nonumber\\
&&\left. - F_{k\lambda \mu }^{\rm{*}}\left( {\hat \theta ,\hat \phi } \right){{\hat b}_{k\lambda \mu }} + F_{k\lambda \mu }^{\rm{*}}\left( {\hat \theta ,\hat \phi } \right){F_{k\lambda \mu }}\left( {\hat \theta ,\hat \phi } \right) \right], \nonumber\\
\end{eqnarray}
and
\begin{eqnarray}
&&{S^{ - 1}}{{\hat H}_{{\mathop{\rm int}} }}S \nonumber\\
&=&\sum\limits_{k\lambda \mu } {{U_\lambda }\left( k \right)} {\rm{ }}\left[ Y_{\lambda \mu }^*\left( {\hat \theta ,\hat \phi } \right)\hat b_{k\lambda \mu }^\dag  - Y_{\lambda \mu }^*\left( {\hat \theta ,\hat \phi } \right)F_{k\lambda \mu }^{\rm{*}}\left( {\hat \theta ,\hat \phi } \right)\right. \nonumber\\
&&\left. + {Y_{\lambda \mu }}\left( {\hat \theta ,\hat \phi } \right){{\hat b}_{k\lambda \mu }} - {Y_{\lambda \mu }}\left( {\hat \theta ,\hat \phi } \right){F_{k\lambda \mu }}\left( {\hat \theta ,\hat \phi } \right) \right].\nonumber\\
\end{eqnarray}

For the derivation of Eq. (8), the operator ${\hat {\bf J}^2}$ is expressed in terms of spherical components ${{\hat J}_i }$ ($i =0, \pm1$) via ${\hat {\bf J}^2} = {{\hat J}_0}^2 + {{\hat J}_0} - 2{{\hat J}_{ - 1}}{{\hat J}_{ + 1}}$\cite{wa4}, then the operator ${{\hat J}_i }$ is transformed by ${{\hat S}^{ - 1}}{{\hat J}_i }\hat S = {{\hat J}_i } + {{\hat P}_i } - {{\hat M}_i }$. The detailed derivations and expressions of ${{\hat P}_i }$ and ${{\hat M}_i }$ are given in Appendix A. In Eqs. (9) and (10), the transformed properties of phonon creation and annihilation operators, $\hat b_{k\lambda \mu }^\dag$ and ${{{\hat b}_{k\lambda \mu }}}$ by the $\hat S$ operator have been used in the following way
\begin{equation}
{{\hat S}^{ - 1}}\sum\limits_{k\lambda \mu } {\hat b_{k\lambda \mu }^\dag } \hat S=\sum\limits_{k\lambda \mu } {\left[ {\hat b_{k\lambda \mu }^\dag  - F_{k\lambda \mu }^{\rm{*}}\left( {\hat \theta ,\hat \phi } \right)} \right]},
\end{equation}
\begin{equation}
{{\hat S}^{ - 1}}\sum\limits_{k\lambda \mu } {{{\hat b}_{k\lambda \mu }}} \hat S = \sum\limits_{k\lambda \mu } {\left[ {{{\hat b}_{k\lambda \mu }} - {F_{k\lambda \mu }}\left( {\hat \theta ,\hat \phi } \right)} \right]}.
\end{equation}
The detailed transformation processes for them are also presented in Appendix A.

We classify $\tilde H$ into three terms including zero-phonon term ${\tilde H_0}$, one-phonon term ${{\tilde H}_1}$ and two-phonon term ${{\tilde H}_2}$, which could be rewritten as
\begin{widetext}
\begin{eqnarray}
{{\tilde H}_0} &=& B{\left( {{{\hat J}_0} - {{\hat M}_0}} \right)^2} + B\left( {{{\hat J}_0} - {{\hat M}_0}} \right) - 2B\left( {{{\hat J}_{ - 1}} - {{\hat M}_{ - 1}}} \right)\left( {{{\hat J}_{ + 1}} - {{\hat M}_{ + 1}}} \right)+\sum\limits_{k\lambda \mu } {{\omega _k}\hat b_{k\lambda \mu }^\dag {{\hat b}_{k\lambda \mu }}}\nonumber\\
&&+ B\sum\limits_{k\lambda \mu } {{{\left[ {{U_\lambda }\left( k \right)} \right]}^2}} {\mu ^2}{Y_{\lambda \mu }}\left( {\hat \theta ,\hat \phi } \right)Y_{\lambda \mu }^*\left( {\hat \theta ,\hat \phi } \right){\mathbb{Z} ^2} + 2B\sum\limits_{k\lambda \mu } {{{\left[ {{U_\lambda }\left( k \right)} \right]}^2}} \frac{{\lambda \left( {\lambda  + 1} \right) - \mu \left( {\mu  - 1} \right)}}{2}{\mathbb{Z} ^2}\nonumber\\
&& + \sum\limits_{k\lambda \mu } {{\omega _k}F_{k\lambda \mu }^{\rm{*}}\left( {\hat \theta ,\hat \phi } \right){F_{k\lambda \mu }}\left( {\hat \theta ,\hat \phi } \right)}  - \sum\limits_{k\lambda \mu } {{U_\lambda }\left( k \right)} \left[ Y_{\lambda \mu }^*\left( {\hat \theta ,\hat \phi } \right)F_{k\lambda \mu }^{\rm{*}}\left( {\hat \theta ,\hat \phi } \right) + {Y_{\lambda \mu }}\left( {\hat \theta ,\hat \phi } \right){F_{k\lambda \mu }}\left( {\hat \theta ,\hat \phi } \right) \right],\nonumber\\
\end{eqnarray}
\end{widetext}
\begin{widetext}
\begin{eqnarray}
 {\tilde H_1} &=& 2B{{\hat P}_0} \left( {{{\hat J}_0} - {{\hat M}_0}} \right) + B{{\hat P}_0} - 2B{{\hat P}_{ - 1}} \left( {{{\hat J}_{ + 1}} - {{\hat M}_{ + 1}}} \right) - 2B{{\hat P}_{ + 1}} \left( {{{\hat J}_{ - 1}} - {{\hat M}_{ - 1}}} \right) \nonumber\\
 &&  + B\sum\limits_{k\lambda \mu } {{U_\lambda }\left( k \right)} {\mu ^2}\left[ {{Y_{\lambda \mu }}\left( {\hat \theta ,\hat \phi } \right){{\hat b}_{k\lambda \mu }} - Y_{\lambda \mu }^ * \left( {\hat \theta ,\hat \phi } \right)\hat b_{k\lambda \mu }^\dag } \right]\mathbb{Z} \nonumber\\
 && + 2B\sum\limits_{k\lambda \mu } {{U_\lambda }\left( k \right)} \left[ {\frac{{\lambda \left( {\lambda  + 1} \right) - \mu \left( {\mu  + 1} \right)}}{2}{Y_{\lambda \mu }}\left( {\hat \theta ,\hat \phi } \right){{\hat b}_{k\lambda \mu }} - \frac{{\lambda \left( {\lambda  + 1} \right) - \mu \left( {\mu  - 1} \right)}}{2}Y_{\lambda \mu }^*\left( {\hat \theta ,\hat \phi } \right)\hat b_{k\lambda \mu }^\dag } \right]\mathbb{Z}  \nonumber\\
 &&  - \sum\limits_{k\lambda \mu } {{\omega _k}\left[ {\hat b_{k\lambda \mu }^\dag {F_{k\lambda \mu }}\left( {\hat \theta ,\hat \phi } \right) + F_{k\lambda \mu }^{\rm{*}}\left( {\hat \theta ,\hat \phi } \right){{\hat b}_{k\lambda \mu }}} \right]}  + \sum\limits_{k\lambda \mu } {{U_\lambda }\left( k \right)} \left[ {Y_{\lambda \mu }^*\left( {\hat \theta ,\hat \phi } \right)\hat b_{k\lambda \mu }^\dag  + {Y_{\lambda \mu }}\left( {\hat \theta ,\hat \phi } \right){{\hat b}_{k\lambda \mu }}} \right],\nonumber\\
\end{eqnarray}
\end{widetext}
\begin{widetext}
\begin{eqnarray}
 {{\tilde H}_2} &=& B\sum\limits_{k\lambda \mu } {{U_\lambda }\left( k \right)} {U_{\lambda '}}\left( {k'} \right)\mu \mu '\left[ {Y_{\lambda \mu }}\left( {\hat \theta ,\hat \phi } \right){Y_{\lambda '\mu '}}\left( {\hat \theta ,\hat \phi } \right){{\hat b}_{k\lambda \mu }}{{\hat b}_{k'\lambda '\mu '}} + {Y_{\lambda \mu }}\left( {\hat \theta ,\hat \phi } \right)Y_{\lambda '\mu '}^ * \left( {\hat \theta ,\hat \phi } \right)\hat b_{k'\lambda '\mu '}^\dag {{\hat b}_{k\lambda \mu }} \right.\nonumber\\
 &&\left.+ Y_{\lambda \mu }^ * \left( {\hat \theta ,\hat \phi } \right){Y_{\lambda '\mu '}}\left( {\hat \theta ,\hat \phi } \right)\hat b_{k\lambda \mu }^\dag {{\hat b}_{k'\lambda '\mu '}} + Y_{\lambda \mu }^ * \left( {\hat \theta ,\hat \phi } \right)Y_{\lambda '\mu '}^ * \left( {\hat \theta ,\hat \phi } \right)\hat b_{k\lambda \mu }^\dag \hat b_{k'\lambda '\mu '}^\dag  \right]{\mathbb{Z} ^2}+ 2B\sum\limits_{k\lambda \mu } {{U_\lambda }\left( k \right)} {U_{\lambda '}}\left( {k'} \right) \nonumber\\
 &&\times \left[ \sqrt {\frac{{\lambda \left( {\lambda  + 1} \right) - \mu \left( {\mu  + 1} \right)}}{2}} {Y_{\lambda \mu  + 1}}\left( {\hat \theta ,\hat \phi } \right)\sqrt {\frac{{\lambda '\left( {\lambda ' + 1} \right) - \mu '\left( {\mu ' - 1} \right)}}{2}} {Y_{\lambda '\mu ' - 1}}\left( {\hat \theta ,\hat \phi } \right){{\hat b}_{k\lambda \mu }}{{\hat b}_{k'\lambda '\mu '}}\right. \nonumber\\
 &&+ \sqrt {\frac{{\lambda \left( {\lambda  + 1} \right) - \mu \left( {\mu  + 1} \right)}}{2}} {Y_{\lambda \mu  + 1}}\left( {\hat \theta ,\hat \phi } \right)\sqrt {\frac{{\lambda '\left( {\lambda ' + 1} \right) - \mu '\left( {\mu ' + 1} \right)}}{2}} Y_{\lambda '\mu ' + 1}^*\left( {\hat \theta ,\hat \phi } \right)\hat b_{k'\lambda '\mu '}^\dag {{\hat b}_{k\lambda \mu }} \nonumber\\
 &&+ \sqrt {\frac{{\lambda \left( {\lambda  + 1} \right) - \mu \left( {\mu  - 1} \right)}}{2}} Y_{\lambda \mu  - 1}^*\left( {\hat \theta ,\hat \phi } \right)\sqrt {\frac{{\lambda '\left( {\lambda ' + 1} \right) - \mu '\left( {\mu ' - 1} \right)}}{2}} {Y_{\lambda '\mu ' - 1}}\left( {\hat \theta ,\hat \phi } \right)\hat b_{k\lambda \mu }^\dag {{\hat b}_{k'\lambda '\mu '}} \nonumber\\
 &&\left.+ \sqrt {\frac{{\lambda \left( {\lambda  + 1} \right) - \mu \left( {\mu  - 1} \right)}}{2}} Y_{\lambda \mu  - 1}^*\left( {\hat \theta ,\hat \phi } \right)\sqrt {\frac{{\lambda '\left( {\lambda ' + 1} \right) - \mu '\left( {\mu ' + 1} \right)}}{2}} Y_{\lambda '\mu ' + 1}^*\left( {\hat \theta ,\hat \phi } \right)\hat b_{k\lambda \mu }^\dag \hat b_{k'\lambda '\mu '}^\dag  \right]{\mathbb{Z} ^2},\nonumber\\
\end{eqnarray}
\end{widetext}
where the one-phonon term ${{\tilde H}_1}$ and two-phonon term ${{\tilde H}_2}$ are neglected due to their minor contribution to the total energy of angulon in the helium environment\cite{wm1,wp1,wxx1}. Thus, we choose the product form of the rotational molecule state $\left| {LM} \right\rangle $ and the zero-phonon state $\left| 0_{\rm ph} \right\rangle $, that is $\left| {LM} \right\rangle \left| 0_{\rm ph} \right\rangle $, as the eigenstate of the angulon. The expectation value of ${{\tilde H}_0}$ is given by
\begin{equation}
E_L^* = \langle \Psi |\tilde H\left| \Psi  \right\rangle  =\left\langle 0_{\rm ph} \right|\langle LM|{\tilde H_0}\left| {LM} \right\rangle \left| 0_{\rm ph} \right\rangle .
\end{equation}
To get the expressions of $\mathbb{Z}$ and $\mathbb{W}$, we carry out the minimization of Eq. (16) with respect to $\mathbb{Z}$ and $\mathbb{W}$
\begin{equation}
\frac{{\partial \left\langle 0_{\rm ph} \right|\langle LM|{{\tilde H}_0}\left| {LM} \right\rangle \left| 0_{\rm ph} \right\rangle }}{{\partial \mathbb{Z}}} = 0,
\end{equation}
\begin{equation}
\frac{{\partial \left\langle 0_{\rm ph} \right|\langle LM|{{\tilde H}_0}\left| {LM} \right\rangle \left| 0_{\rm ph} \right\rangle }}{{\partial \mathbb{W} }} = 0.
\end{equation}
The detailed variational calculations are given in Appendix B. In most experiments, the evaluation of the effective value for $B^*$ is usually determined from the splitting between the two lowest rotational states\cite{w8,wm1}. For the sake of simplicity, we select the first-excited rotational state $\left| {LM} \right\rangle = \left| {10} \right\rangle$ in Eq. (16), which suffices to estimate the renormalization of the rotational constants for different molecular species as shown in the following sections. Meanwhile, the dominate channel of the phonon angular momentum $\lambda=2$, namely, the quantum state $\left| {\lambda \mu} \right\rangle = \left| {20} \right\rangle$ is considered for the coupling between molecule and helium in most cases\cite{w1,wm1,wp1,wxx1}. After the complicated calculations for Eqs. (16), (17) and (18), one can get
\begin{equation}
\mathbb{Z}  = \frac{{\alpha  - {\chi ^2}}}{{6B\gamma  + {\omega _k}\left( {\alpha  - {\chi ^2}} \right)}},
\end{equation}
and
\begin{equation}
\mathbb{W}  = \frac{{6B\gamma \chi }}{{{\omega _k}\left[ {6B\gamma  + {\omega _k}\left( {\alpha  - {\chi ^2}} \right)} \right]}},
\end{equation}
\nonumber\\
with $\chi  = \langle 10|{Y_{20}}\left| {10} \right\rangle$, $\alpha  = \langle 10|{\left| {{Y_{20}}} \right|^2}\left| {10} \right\rangle$, and $\gamma  = \langle 10|{\left| {{Y_{21}}} \right|^2}\left| {10} \right\rangle$, respectively.

Substituting Eqs. (19) and (20) into Eq. (16), one can obtain the angulon energy for the first-excited rotational state with an energy shift
\begin{align}
{E^ * } =& 2B + {\sum\limits_{k} {\left[ {{U_2}\left( k \right)} \right]} ^2}\left[ \left( {6B\gamma  + {\omega _k}} \right){\mathbb{Z} ^2} + {\omega _k}{\mathbb{W} ^2} \right.\nonumber\\
&\left.+ 2\chi {\omega _k}\mathbb{Z} \mathbb{W}  - 2\alpha \mathbb{Z}  - 2\chi \mathbb{W} \right]\nonumber\\
=& 2B - {\sum\limits_k {\left[ {{U_2}\left( k \right)} \right]} ^2} \left( \alpha \mathbb{Z}  + \chi \mathbb{W} \right ),\tag{21}
\end{align}
casting this energy shift into the renormalization of the rotational constant via ${E^ * } = 2{B^ * }$, the ratio can be written as
\begin{align}
\frac{{{B^ * }}}{B}=&1 - \sum\limits_k {\frac{{{{\left[ {{U_\lambda }\left( k \right)} \right]}^2}}}{{2\pi B}}\left[ {\frac{{11}}{{28}}\frac{1}{{10B + {\omega _k}}} + \frac{1}{{\left( {10B + {\omega _k}} \right){\omega _k}}}} \right]}.\tag{22}
\end{align}

In the numerical simulation, these adopted parameters for the superfluid helium are given as follows\cite{w1,w2,wf1,wf2,wf3,w34,wm1,wp1}: ${g_{bb}} = 418{\left( {{m^3}{u_0}} \right)^{ - {1/2}} }$, $n = 0.014{\left( {m{u_0}} \right)^{3/2}}$ with ${u_0}=218B$ and $m=1.63\times10^ {-25}{\rm {kg}}$; the cut-off wave vector of the phonon bath $k_c=5/{\rm {\AA}} $.
\begin{figure}
\includegraphics[width=3.7in,keepaspectratio]{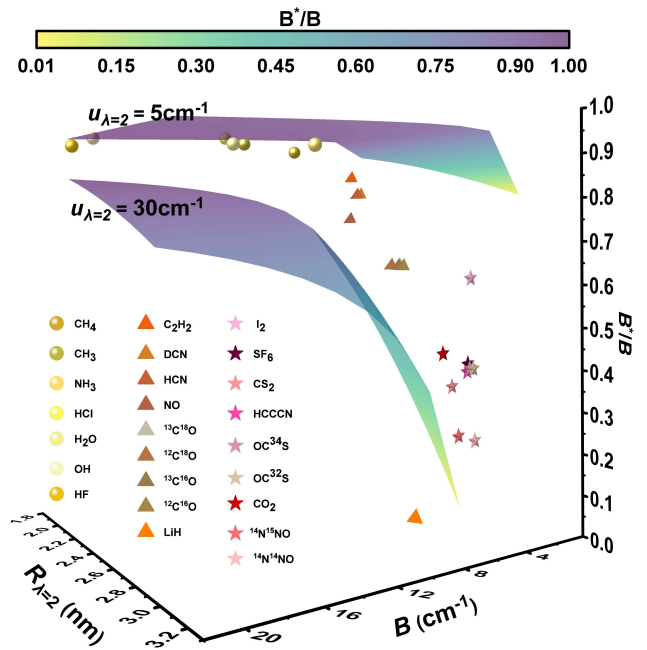}
\caption{\label{compare} The renormalization of the rotational constant ${B^*}/B$ as functions of the intrinsic rotational constant $B$ and the effective radius of the solvation shell for different kinds of molecules for the coupling strength ${u_{\lambda=2}}$ in the range of ${u_{\lambda=2}}=5\sim30$ cm$^{-1}$, where the dominant channel of the phonon angular momentum ${\lambda=2}$ is considered. Experimental data for these molecules are from Refs.\cite{w25,w27,w37,w38,w39,w40,w41,w42,w43,w44,w45,w46,w47,w48,w49,w50,w51,w52,w53,w54}, which are classified into the weak-coupling species (circles), the intermediate-coupling species (triangles) and the strong-coupling species(pentalphas).}
\end{figure}
\section{Result and Discussion}
According to Eq. (22), we present the effective renormalization of rotational constants for different molecules for the strength of molecule-helium coupling between ${u_{\lambda=2}}=5$ and $30$ cm$^{-1}$ in Fig. 1. One can see that a very good agreement with the collected experiment data\cite{w25,w27,w37,w38,w39,w40,w41,w42,w43,w44,w45,w46,w47,w48,w49,w50,w51,w52,w53,w54} for three species: light-mass (circles), intermediate-mass (triangles) and heavy-mass molecules (pentalphas). These results show that this all-coupling model (i) fills up the gap between the strong- and weak-coupling limits, where only a rough estimation by interpolating between the weak- and strong-coupling theories was carried out for the intermediate-coupling; (ii) could avoid these overestimations or underestimations of the renormalization effect for these molecules in previous studies.  Another advantage of this model is that the effective radius of the helium solvation shell corotating along with the molecules could be evaluated for these three species. We see that the radius increases obviously from the light-mass molecules to the intermediated-and the heavy-mass ones, which means that the bigger radius of the solvation shell just corresponds to the stronger strength of the helium-molecules coupling, leading to the larger renormalization of rotational constant. In order to show these effect clearly, we further list three typical molecular species in Table 1 extracted from Fig. 1. This effective radius not only describes properly the scale of the coupling potential between molecules and superfluid helium, but is more directly related to some measurable parameters in experiments, e.g., the centrifugal distortion constants of superfluid helium droplets\cite{wx2}.
\begin{table}[htbp]
\renewcommand\arraystretch{1.5}
\caption{\label{compare} The parameters of three typical molecules CH$_4$, NO and ${^{14}}$N${^{14}}$NO extracted from the Fig. 1, which correspond to the weak-, intermediate- and strong-coupling cases, respectively.}
\centering
	\begin{tabular}{|c|c|c|c|c|}\hline
\makebox[0.1\textwidth][c]{Molecules}
		  &B(cm$^{-1}$)&${B^ *} /B$& $u_{\lambda=2}$(cm$^{-1}$) &$R_{\lambda=2}$(nm) \\ \hline
		CH$_4$ &5.25&0.96&6.7&1.92  \\
        NO &1.65 &0.76 &17.82 &2.36  \\
        ${^{14}}$N${^{14}}$NO     &0.42 &0.17&19.46 & 2.57  \\ \hline
	\end{tabular}
\end{table}

It is worth noting that for LiH molecule in Fig. 1, which belongs to the intermediate-mass species, but with the strong renormalization of the rotational constant proved by experiments\cite{wx2,w55} and predicted by the quantum Monte Carlo calculation\cite{wm1,w42}. Based on the strong-coupling angulon theory, Lemeshko attributed it to the fact that the pronounced anisotropy of the helium-LiH potential energy surface, where the contribution of the other channels of phonon angular momentum, such as ${u_{\lambda=1}}$\cite{wm1,w55,wttt0,wttt1}, should be considered. In order to reproduce this strong renormalization prediction of LiH, the values for parameters ${u_{\lambda=2}}=37$ cm$^{-1}$ and ${R_{\lambda=2}}=3.03$ nm  could be adopted in the present model. This implies that these appropriate values for the strength of molecule-helium coupling and the effective radius of the solvation shell could be evaluated by this model to fitting the experimental measurements, showing the widely applicability of this all-coupling theory.

Here, we must emphasize that (i) the magnitude of the renormalization decreases for the larger rotational quantum state, such as $L=2, 3\cdots$, corresponding to the higher rotational speed of molecules. The reason is that the surrounding phononic bath will be not able to follow the rotational motion when molecules rotating faster and faster, even decoupling from them at certain rotational speed (state), which have been proved by recent experiments\cite{wm1,wp1,w55}; (ii) apart from the mainly channel of the phonon angular momentum (${u_{\lambda=2}}$), other channels, in fact, could also be added by following the same processes given in this model if their contribution are not neglected; (ii) the present model is not designed to compete with first-principles approaches in accuracy, such as the path-integral or quantum Monte Carlo calculations\cite{w40,w41,w56,w57}. However, this very simple expression of Eq. (22) could effectively fit the collected experimental data in the broad range of molecular species as shown above. We hope this model provides the qualitative predictions for particles rotating in quantum many-body environment along with simple explanations for the underlying physics and stimulates further experiments.

In summary, we have successfully reproduced the effective rotational constants for different molecules rotating in superfluid helium by introducing an all-coupling angulon model, where the strength of molecule-helium coupling and the effective radius of the solvation shell corotating along with the molecular rotor could be evaluated qualitatively. These results not only provide the significant enlightenment for studying the molecular dynamics in the phononic environment, but also open a new way for systematically analyzing phonon angular momenta transfer in angulon frame.

\section*{ACKNOWLEDGMENT}
This work was supported by National Natural Science Foundation of China (Nos 11674241 and 12174283).

\begin{widetext}
\begin{appendix}
\section{The detailed derivations for the transformation of angulon hamiltonian}
Here, we provide the derivations for the transformed Hamiltonian in detail.

Firstly, the relations of
\begin{eqnarray}
{Y_{\lambda \mu }}\left( {\hat \theta ,\hat \phi } \right) = \sqrt {\frac{{2\lambda  + 1}}{{4\pi }}} D_{\mu 0}^{\lambda  * }\left( {\hat \phi ,\hat \theta ,0} \right),
\end{eqnarray}
and
\begin{eqnarray}
Y_{\lambda \mu }^ * \left( {\hat \theta ,\hat \phi } \right) = \sqrt {\frac{{2\lambda  + 1}}{{4\pi }}} D_{\mu 0}^\lambda \left( {\hat \phi ,\hat \theta ,0} \right),
\end{eqnarray}
are employed, where, $D_{\mu 0}^\lambda \left( {\hat \phi ,\hat \theta ,0} \right)$ are Wigner $D$ matrices whose arguments are the angle operators defining the orientation of the molecule\cite{wp1}.

Secondly, the operators ${{\hat J}_i}$ ($i = \pm1,0$) obeys the following commutation relations with Wigner $D$ matrix
\begin{eqnarray}
\left[ {{{\hat J}_i},D_{\mu v}^\lambda \left( {\hat \phi ,\hat \theta ,\hat \gamma } \right)} \right] = {\left( { - 1} \right)^{i + 1}}\sqrt {\lambda \left( {\lambda  + 1} \right)} C_{\lambda ,\mu ;1, - i}^{\lambda ,\mu  - i}D_{\mu  - i,v}^\lambda \left( {\hat \phi ,\hat \theta ,\hat \gamma } \right),
\end{eqnarray}
\begin{eqnarray}
\left[ {{{\hat J}_i},D_{\mu v}^{\lambda  * }\left( {\hat \phi ,\hat \theta ,\hat \gamma } \right)} \right] = \sqrt {\lambda \left( {\lambda  + 1} \right)} C_{\lambda ,\mu ;1,i}^{\lambda ,\mu  + i}D_{\mu  + i,v}^{\lambda  * } \left( {\hat \phi ,\hat \theta ,\hat \gamma } \right),
\end{eqnarray}
where $C_{{l_1},{m_1};{l_2},{m_2}}^{{l_3},{m_3}}$ are the Clebsch-Gordan (C-G) coefficients.

Thirdly, the Taylor expansion is introduced for the operator
\begin{eqnarray}
exp\left[ { - V} \right]{\rm{\;}}a{\rm{\;}}exp\left[ V \right]= a + \left[ {a,V} \right] + \frac{1}{2}\left[ {\left[ {a,V} \right],V} \right] + \frac{1}{{3!}}\left[ {\left[ {\left[ {a,V} \right],V} \right],V} \right] +  \cdot  \cdot  \cdot.
\end{eqnarray}
Based on these relations, one can get
\begin{align}
&{{\hat S}^{ - 1}}{{\hat J}_0}\hat S \nonumber\\
=&{{\hat J}_0} + \sum\limits_{k\lambda \mu } {{U_\lambda }\left( k \right)\mathbb{Z}\sqrt {\frac{{2\lambda  + 1}}{{4\pi }}} } \sqrt {\lambda \left( {\lambda  + 1} \right)} \left[ {C_{\lambda ,\mu ;1,0}^{\lambda ,\mu }D_{\mu ,0}^{\lambda *}\left( {\hat \phi ,\hat \theta ,0} \right){{\hat b}_{k\lambda \mu }} + C_{\lambda ,\mu ;1,0}^{\lambda ,\mu }D_{\mu ,0}^\lambda \left( {\hat \phi ,\hat \theta ,0} \right)\hat b_{k\lambda \mu }^\dag } \right] \nonumber\\
&- \frac{1}{2}\sum\limits_{k\lambda \mu } {{U_\lambda }} \left( k \right)\mathbb{Z}\sqrt {\frac{{2\lambda  + 1}}{{4\pi }}} \sqrt {\lambda \left( {\lambda  + 1} \right)} \{ C_{\lambda ,\mu ;1,0}^{\lambda ,\mu }D_{\mu ,0}^{\lambda *}\left( {\hat \phi ,\hat \theta ,0} \right)\left[ {{U_\lambda }\left( k \right)\mathbb{Z} \sqrt {\frac{{2\lambda  + 1}}{{4\pi }}} D_{\mu ,0}^\lambda \left( {\hat \phi ,\hat \theta ,0} \right) + {U_\lambda }\left( k \right)\mathbb{W}} \right] \nonumber\\
&+ C_{\lambda ,\mu ;1,0}^{\lambda ,\mu }D_{\mu ,0}^\lambda \left( {\hat \phi ,\hat \theta ,0} \right)\left[ {{U_\lambda }\left( k \right)\mathbb{Z} \sqrt {\frac{{2\lambda  + 1}}{{4\pi }}} D_{\mu ,0}^{\lambda *}\left( {\hat \phi ,\hat \theta ,0} \right) + {U_\lambda }\left( k \right)\mathbb{W}} \right]\}  \nonumber\\
=&{{\hat J}_0} + \sum\limits_{k\lambda \mu } {{U_\lambda }\left( k \right)\mu } \left[ {{Y_{\lambda \mu }}\left( {\hat \theta ,\hat \phi } \right){{\hat b}_{k\lambda \mu }} + Y_{\lambda \mu }^ * \left( {\hat \theta ,\hat \phi } \right)\hat b_{k\lambda \mu }^\dag } \right]\mathbb{Z} \nonumber\\
&- \frac{1}{2}\sum\limits_{k\lambda \mu } {{{\left[ {{U_\lambda }\left( k \right)} \right]}^2}} \mu \left\{ {\left[ {{Y_{\lambda \mu }}\left( {\hat \theta ,\hat \phi } \right)Y_{\lambda \mu }^ * \left( {\hat \theta ,\hat \phi } \right)\mathbb{Z}  + {Y_{\lambda \mu }}\left( {\hat \theta ,\hat \phi } \right)\mathbb{W} } \right] + \left[ {Y_{\lambda \mu }^ * \left( {\hat \theta ,\hat \phi } \right){Y_{\lambda \mu }}\left( {\hat \theta ,\hat \phi } \right)\mathbb{Z}  + Y_{\lambda \mu }^ * \left( {\hat \theta ,\hat \phi } \right)\mathbb{W} } \right]} \right\}\mathbb{Z}\nonumber\\
=&{{\hat J}_0} +{\hat P_0}-{{\hat M}_0}, \tag{A6}
\end{align}
with the abbreviations
\begin{align}
{\hat P_0}=\sum\limits_{k\lambda \mu } {{U_\lambda }\left( k \right)\mu } \left[ {{Y_{\lambda \mu }}\left( {\hat \theta ,\hat \phi } \right){{\hat b}_{k\lambda \mu }} + Y_{\lambda \mu }^*\left( {\hat \theta ,\hat \phi } \right)\hat b_{k\lambda \mu }^\dag } \right]\mathbb{\mathbb{Z}},
\end{align}
\begin{align}
{{\hat M}_0} =  \frac{1}{2}\sum\limits_{k\lambda \mu } {{{\left[ {{U_\lambda }\left( k \right)} \right]}^2}\mu } \{ \left[ {{Y_{\lambda \mu }}\left( {\hat \theta ,\hat \phi } \right)Y_{\lambda \mu }^*\left( {\hat \theta ,\hat \phi } \right)\mathbb{Z}  + {Y_{\lambda \mu }}\left( {\hat \theta ,\hat \phi } \right)\mathbb{W} } \right] + \left[ {Y_{\lambda \mu }^*\left( {\hat \theta ,\hat \phi } \right){Y_{\lambda \mu }}\left( {\hat \theta ,\hat \phi } \right)\mathbb{Z}  + Y_{\lambda \mu }^*\left( {\hat \theta ,\hat \phi } \right)\mathbb{W} } \right]\} \mathbb{Z}.
\end{align}
Similarly, the components of the transformation for the rest two terms ${{\hat J}_{-1}}$ and ${{\hat J}_{+1}}$ can be written as
\begin{align}
{\hat S^{ - 1}}{\hat J_{ - 1}}\hat S ={\hat J_{-1}} + {\hat P_{-1}} - {\hat M_{-1}},\tag{A7}
\end{align}
with
\begin{align}
{\hat P_{-1}}=\sum\limits_{k\lambda \mu } {{U_\lambda }} \left( k \right) [\sqrt {\frac{{\lambda \left( {\lambda  + 1} \right) - \mu \left( {\mu  - 1} \right)}}{{2}}} {Y_{\lambda \mu  - 1}}\left( {\hat \theta ,\hat \phi } \right){{\hat b}_{k\lambda \mu }}+ \sqrt {\frac{{\lambda \left( {\lambda  + 1} \right) - \mu \left( {\mu  + 1} \right)}}{{2}}} Y_{\lambda \mu  + 1}^*\left( {\hat \theta ,\hat \phi } \right)\hat b_{k\lambda \mu }^\dag ] \mathbb{Z},
\end{align}
\begin{align}
{\hat M_{-1}}=&  \frac{1}{2}\sum\limits_{k\lambda \mu } {{{\left[ {{U_\lambda }\left( k \right)} \right]}^2}} \left\{ \sqrt {\frac{{\lambda \left( {\lambda  + 1} \right) - \mu \left( {\mu  - 1} \right)}}{2}} \left[ {{Y_{\lambda \mu  - 1}}\left( {\hat \theta ,\hat \phi } \right)\mathbb{W}  + {Y_{\lambda \mu  - 1}}\left( {\hat \theta ,\hat \phi } \right)Y_{\lambda \mu }^*\left( {\hat \theta ,\hat \phi } \right)\mathbb{Z} } \right]\right.    \nonumber\\
& \left.+ \sqrt {\frac{{\lambda \left( {\lambda  + 1} \right) - \mu \left( {\mu  + 1} \right)}}{2}} \left[ {Y_{\lambda \mu  + 1}^*\left( {\hat \theta ,\hat \phi } \right)\mathbb{W}  + Y_{\lambda \mu  + 1}^*\left( {\hat \theta ,\hat \phi } \right){Y_{\lambda \mu }}\left( {\hat \theta ,\hat \phi } \right)\mathbb{Z} } \right] \right\}\mathbb{Z},
\end{align}
and
\begin{align}
{\hat S^{ - 1}}{\hat J_{ + 1}}\hat S ={\hat J_{+1}} + {\hat P_{+1}} - {\hat M_{+1}},\tag{A8}
\end{align}
with
\begin{align}
{\hat P_{+1}}= - \sum\limits_{k\lambda \mu } {{U_\lambda }\left( k \right)} [\sqrt {\frac{{\lambda \left( {\lambda  + 1} \right) - \mu \left( {\mu  + 1} \right)}}{2}} {Y_{\lambda \mu  + 1}}\left( {\hat \theta ,\hat \phi } \right){{\hat b}_{k\lambda \mu }} + \sqrt {\frac{{\lambda \left( {\lambda  + 1} \right) - \mu \left( {\mu  - 1} \right)}}{2}} Y_{\lambda \mu  - 1}^*\left( {\hat \theta ,\hat \phi } \right)\hat b_{k\lambda \mu }^\dag ]\mathbb{Z},
\end{align}
\begin{align}
{\hat M_{+1}}=& - \frac{1}{2}\sum\limits_{k\lambda \mu } {{{\left[ {{U_\lambda }\left( k \right)} \right]}^2}} \left\{ \sqrt {\frac{{\lambda \left( {\lambda  + 1} \right) - \mu \left( {\mu  + 1} \right)}}{2}} \left[ {{Y_{\lambda \mu  + 1}}\left( {\hat \theta ,\hat \phi } \right)\mathbb{W}  + {Y_{\lambda \mu  + 1}}\left( {\hat \theta ,\hat \phi } \right)Y_{\lambda \mu }^*\left( {\hat \theta ,\hat \phi } \right)\mathbb{Z} } \right] \right.  \nonumber\\
 & \left. + \sqrt {\frac{{\lambda \left( {\lambda  + 1} \right) - \mu \left( {\mu  - 1} \right)}}{2}} \left[ {Y_{\lambda \mu  - 1}^*\left( {\hat \theta ,\hat \phi } \right)\mathbb{W}  + Y_{\lambda \mu  - 1}^*\left( {\hat \theta ,\hat \phi } \right){Y_{\lambda \mu }}\left( {\hat \theta ,\hat \phi } \right)\mathbb{Z} } \right] \right\}\mathbb{Z}.
\end{align}

According to the derivations for ${{\hat P}_0}$, ${{\hat P}_{-1}}$ and ${{\hat P}_{+1}}$, we get
\begin{align}
{{\hat P}_0}^2 =&\sum\limits_{k\lambda \mu } {{U_\lambda }\left( k \right)} {U_{\lambda '}}\left( {k'} \right)\mu \mu '{\mathbb{Z} ^2} \nonumber\\
&\times\left[ {Y_{\lambda \mu }}\left( {\hat \theta ,\hat \phi } \right){Y_{\lambda '\mu '}}\left( {\hat \theta ,\hat \phi } \right){{\hat b}_{k\lambda \mu }}{{\hat b}_{k'\lambda '\mu '}} + {Y_{\lambda \mu }}\left( {\hat \theta ,\hat \phi } \right)Y_{\lambda '\mu '}^ * \left( {\hat \theta ,\hat \phi } \right)\left( {\hat b_{k'\lambda '\mu '}^\dag {{\hat b}_{k\lambda \mu }} + {\delta _{kk'}}{\delta _{\lambda \lambda '}}{\delta _{\mu \mu '}}} \right) \right. \nonumber\\
&\left. + Y_{\lambda \mu }^ * \left( {\hat \theta ,\hat \phi } \right){Y_{\lambda '\mu '}}\left( {\hat \theta ,\hat \phi } \right)\hat b_{k\lambda \mu }^\dag {{\hat b}_{k'\lambda '\mu '}} + Y_{\lambda \mu }^ * \left( {\hat \theta ,\hat \phi } \right)Y_{\lambda '\mu '}^ * \left( {\hat \theta ,\hat \phi } \right)\hat b_{k\lambda \mu }^\dag \hat b_{k'\lambda '\mu '}^\dag \right],\tag{A9}
\end{align}
\begin{align}
&{{\hat P}_{ - 1}} {{\hat P}_{ + 1}}\nonumber\\
=&\sum\limits_{k\lambda \mu } {{U_\lambda }\left( k \right)} {U_{\lambda '}}\left( {k'} \right){\mathbb{Z} ^2}\nonumber\\
&\times \left[ \sqrt {\frac{{\lambda \left( {\lambda  + 1} \right) - \mu \left( {\mu  + 1} \right)}}{2}} {Y_{\lambda \mu  + 1}}\left( {\hat \theta ,\hat \phi } \right)\sqrt {\frac{{\lambda '\left( {\lambda ' + 1} \right) - \mu '\left( {\mu ' - 1} \right)}}{2}} {Y_{\lambda '\mu ' - 1}}\left( {\hat \theta ,\hat \phi } \right){{\hat b}_{k\lambda \mu }}{{\hat b}_{k'\lambda '\mu '}}\right.\nonumber\\
& + \sqrt {\frac{{\lambda \left( {\lambda  + 1} \right) - \mu \left( {\mu  + 1} \right)}}{2}} {Y_{\lambda \mu  + 1}}\left( {\hat \theta ,\hat \phi } \right)\sqrt {\frac{{\lambda '\left( {\lambda ' + 1} \right) - \mu '\left( {\mu ' + 1} \right)}}{2}} Y_{\lambda '\mu ' + 1}^*\left( {\hat \theta ,\hat \phi } \right)\left( {\hat b_{k'\lambda '\mu '}^\dag {{\hat b}_{k\lambda \mu }} + {\delta _{kk'}}{\delta _{\lambda \lambda '}}{\delta _{\mu \mu '}}} \right) \nonumber\\
& + \sqrt {\frac{{\lambda \left( {\lambda  + 1} \right) - \mu \left( {\mu  - 1} \right)}}{2}} Y_{\lambda \mu  - 1}^*\left( {\hat \theta ,\hat \phi } \right)\sqrt {\frac{{\lambda '\left( {\lambda ' + 1} \right) - \mu '\left( {\mu ' - 1} \right)}}{2}} {Y_{\lambda '\mu ' - 1}}\left( {\hat \theta ,\hat \phi } \right)\hat b_{k\lambda \mu }^\dag {{\hat b}_{k'\lambda '\mu '}} \nonumber\\
&\left.+ \sqrt {\frac{{\lambda \left( {\lambda  + 1} \right) - \mu \left( {\mu  - 1} \right)}}{2}} Y_{\lambda \mu  - 1}^*\left( {\hat \theta ,\hat \phi } \right)\sqrt {\frac{{\lambda '\left( {\lambda ' + 1} \right) - \mu '\left( {\mu ' + 1} \right)}}{2}} Y_{\lambda '\mu ' + 1}^*\left( {\hat \theta ,\hat \phi } \right)\hat b_{k\lambda \mu }^\dag \hat b_{k'\lambda '\mu '}^\dag \right], \tag{A10}
\end{align}
which are merged into the transformed terms ${{\tilde H}_0}$ and ${{\tilde H}_2}$.

In virtue of Eq. (A5), the transformation process of creation operator $\hat b_{k\lambda \mu }^\dag$ is given as
\begin{align}
& {{\hat S}^{ - 1}}\sum\limits_{k\lambda \mu } {\hat b_{k\lambda \mu }^\dag } \hat S \nonumber\\
=& exp\left[ { - \sum\limits_{k\lambda \mu } {[F_{k\lambda \mu }^{\rm{*}}\left( {\hat \theta ,\hat \phi } \right){{\hat b}_{k\lambda \mu }} - {F_{k\lambda \mu }}\left( {\hat \theta ,\hat \phi } \right)\hat b_{k\lambda \mu }^\dag ]} } \right]\sum\limits_{k\lambda \mu } {\hat b_{k\lambda \mu }^\dag } exp\left[ {\sum\limits_{k\lambda \mu } {[F_{k\lambda \mu }^{\rm{*}}\left( {\hat \theta ,\hat \phi } \right){{\hat b}_{k\lambda \mu }} - {F_{k\lambda \mu }}\left( {\hat \theta ,\hat \phi } \right)\hat b_{k\lambda \mu }^\dag ]} } \right]\nonumber\\
=& \sum\limits_{k\lambda \mu } {\hat b_{k\lambda \mu }^\dag }  + \left[ {\hat b_{k\lambda \mu }^\dag ,\sum\limits_{k\lambda \mu } {[F_{k\lambda \mu }^{\rm{*}}\left( {\hat \theta ,\hat \phi } \right){{\hat b}_{k\lambda \mu }} - {F_{k\lambda \mu }}\left( {\hat \theta ,\hat \phi } \right)\hat b_{k\lambda \mu }^\dag ]} } \right] \nonumber\\
=& \sum\limits_{k\lambda \mu } {\hat b_{k\lambda \mu }^\dag }  + \sum\limits_{k\lambda \mu } {\left[ {F_{k\lambda \mu }^{\rm{*}}\left( {\hat \theta ,\hat \phi } \right)\hat b_{k\lambda \mu }^\dag {{\hat b}_{k\lambda \mu }} - {F_{k\lambda \mu }}\left( {\hat \theta ,\hat \phi } \right)\hat b_{k\lambda \mu }^\dag \hat b_{k\lambda \mu }^\dag  - F_{k\lambda \mu }^{\rm{*}}\left( {\hat \theta ,\hat \phi } \right){{\hat b}_{k\lambda \mu }}\hat b_{k\lambda \mu }^\dag  + {F_{k\lambda \mu }}\left( {\hat \theta ,\hat \phi } \right)\hat b_{k\lambda \mu }^\dag \hat b_{k\lambda \mu }^\dag } \right]}  \nonumber\\
=& \sum\limits_{k\lambda \mu } {\left[ {\hat b_{k\lambda \mu }^\dag  - F_{k\lambda \mu }^{\rm{*}}\left( {\hat \theta ,\hat \phi } \right)} \right]}.\tag{A11}
\end{align}
Following the similar process, the transformed annihilation operator ${{\hat b}_{k\lambda \mu }}$ is written as
\begin{align}
{{\hat S}^{ - 1}}\sum\limits_{k\lambda \mu } {{{\hat b}_{k\lambda \mu }}} \hat S=\sum\limits_{k\lambda \mu } {\left[ {{{\hat b}_{k\lambda \mu }} - {F_{k\lambda \mu }}\left( {\hat \theta ,\hat \phi } \right)} \right]}.\tag{A12}
\end{align}
\newpage
\begin{figure}[t]
\section{The variational calculations for the parameters $\mathbb{Z}$ AND $\mathbb{W}$  }
\begin{flushleft}
  From Eqs. (17) and (18), we obtain
\end{flushleft}
\begin{align}
&\frac{{\partial \left\langle 0_{\rm ph} \right|\langle LM|{{\tilde H}_0}\left| {LM} \right\rangle \left| 0_{\rm ph} \right\rangle }}{{\partial \mathbb{Z}}}
= \left\langle 0_{\rm ph} \right|\left\langle {LM\left| {\frac{{\partial {{\tilde H}_0}}}{{\partial \mathbb{Z}}}} \right|LM} \right\rangle \left| 0_{\rm ph} \right\rangle  \nonumber\\
=&\langle LM|B\sum\limits_{k\lambda \mu } {{{\left[ {{U_\lambda }\left( k \right)} \right]}^2}} \mu \left\{ {\left[ {{Y_{\lambda \mu }}\left( {\hat \theta ,\hat \phi } \right)Y_{\lambda \mu }^*\left( {\hat \theta ,\hat \phi } \right) + Y_{\lambda \mu }^*\left( {\hat \theta ,\hat \phi } \right){Y_{\lambda \mu }}\left( {\hat \theta ,\hat \phi } \right)} \right]2\mathbb{Z}  + \left[ {{Y_{\lambda \mu }}\left( {\hat \theta ,\hat \phi } \right) + Y_{\lambda \mu }^*\left( {\hat \theta ,\hat \phi } \right)} \right]\mathbb{W} } \right\}\left( {{{\hat M}_0} - {{\hat J}_0}} \right) \nonumber\\
& - \frac{B}{2}\sum\limits_{k\lambda \mu } {{{\left[ {{U_\lambda }\left( k \right)} \right]}^2}} {\mu ^2}\left[ {{Y_{\lambda \mu }}\left( {\hat \theta ,\hat \phi } \right) - Y_{\lambda \mu }^*\left( {\hat \theta ,\hat \phi } \right)} \right] \mathbb{W} \nonumber\\
&- \frac{B}{2}\sum\limits_{k\lambda \mu } {{{\left[ {{U_\lambda }\left( k \right)} \right]}^2}} \mu \left\{ {\left[ {{Y_{\lambda \mu }}\left( {\hat \theta ,\hat \phi } \right)Y_{\lambda \mu }^*\left( {\hat \theta ,\hat \phi } \right) + Y_{\lambda \mu }^*\left( {\hat \theta ,\hat \phi } \right){Y_{\lambda \mu }}\left( {\hat \theta ,\hat \phi } \right)} \right]2\mathbb{Z}  + \left[ {{Y_{\lambda \mu }}\left( {\hat \theta ,\hat \phi } \right) + Y_{\lambda \mu }^*\left( {\hat \theta ,\hat \phi } \right)} \right]\mathbb{W} } \right\}\nonumber\\
&- \frac{B}{2}\sum\limits_{k\lambda \mu } {{{\left[ {{U_\lambda }\left( k \right)} \right]}^2}} \left( {{{\hat M}_{ + 1}} - {{\hat J}_{ + 1}}} \right)\left\{ \left[ {\sqrt {\frac{{\lambda \left( {\lambda  + 1} \right) - \mu \left( {\mu  - 1} \right)}}{2}} {Y_{\lambda \mu  - 1}}\left( {\hat \theta ,\hat \phi } \right) + \sqrt {\frac{{\lambda \left( {\lambda  + 1} \right) - \mu \left( {\mu  + 1} \right)}}{2}} Y_{\lambda \mu  + 1}^*\left( {\hat \theta ,\hat \phi } \right)} \right]\mathbb{W}  \right. \nonumber\\
&\left.+ \left[ {\sqrt {\frac{{\lambda \left( {\lambda  + 1} \right) - \mu \left( {\mu  - 1} \right)}}{2}} {Y_{\lambda \mu  - 1}}\left( {\hat \theta ,\hat \phi } \right)Y_{\lambda \mu }^*\left( {\hat \theta ,\hat \phi } \right) + \sqrt {\frac{{\lambda \left( {\lambda  + 1} \right) - \mu \left( {\mu  + 1} \right)}}{2}} Y_{\lambda \mu  + 1}^*\left( {\hat \theta ,\hat \phi } \right){Y_{\lambda \mu }}\left( {\hat \theta ,\hat \phi } \right)} \right]2\mathbb{Z}  \right\}\nonumber\\
&  + B\sum\limits_{k\lambda \mu } {{{\left[ {{U_\lambda }\left( k \right)} \right]}^2}} \left( {{{\hat M}_{ - 1}} - {{\hat J}_{ - 1}}} \right)\left\{  \left[ {\sqrt {\frac{{\lambda \left( {\lambda  + 1} \right) - \mu \left( {\mu  + 1} \right)}}{2}} {Y_{\lambda \mu  + 1}}\left( {\hat \theta ,\hat \phi } \right) + \sqrt {\frac{{\lambda \left( {\lambda  + 1} \right) - \mu \left( {\mu  - 1} \right)}}{2}} Y_{\lambda \mu  - 1}^*\left( {\hat \theta ,\hat \phi } \right)} \right]\mathbb{W}  \right. \nonumber\\
&\left. + \left[ {\sqrt {\frac{{\lambda \left( {\lambda  + 1} \right) - \mu \left( {\mu  + 1} \right)}}{2}} {Y_{\lambda \mu  + 1}}\left( {\hat \theta ,\hat \phi } \right)Y_{\lambda \mu }^*\left( {\hat \theta ,\hat \phi } \right) + \sqrt {\frac{{\lambda \left( {\lambda  + 1} \right) - \mu \left( {\mu  - 1} \right)}}{2}} Y_{\lambda \mu  - 1}^*\left( {\hat \theta ,\hat \phi } \right){Y_{\lambda \mu }}\left( {\hat \theta ,\hat \phi } \right)} \right]2\mathbb{Z} \right\}\nonumber\\
&+B\sum\limits_{k\lambda \mu } {{{\left[ {{U_\lambda }\left( k \right)} \right]}^2}} \left\{ \frac{{\lambda \left( {\lambda  + 1} \right) - \mu \left( {\mu  + 1} \right)}}{2}\left[ {{Y_{\lambda \mu }}\left( {\hat \theta ,\hat \phi } \right)Y_{\lambda \mu }^*\left( {\hat \theta ,\hat \phi } \right) - {Y_{\lambda \mu  + 1}}\left( {\hat \theta ,\hat \phi } \right)Y_{\lambda \mu  + 1}^*\left( {\hat \theta ,\hat \phi } \right)} \right]2\mathbb{Z}\right.  \nonumber\\
&  - \frac{{\lambda \left( {\lambda  + 1} \right) - \mu \left( {\mu  - 1} \right)}}{2}\left[ {Y_{\lambda \mu }^*\left( {\hat \theta ,\hat \phi } \right){Y_{\lambda \mu }}\left( {\hat \theta ,\hat \phi } \right) - Y_{\lambda \mu  - 1}^*\left( {\hat \theta ,\hat \phi } \right){Y_{\lambda \mu  - 1}}\left( {\hat \theta ,\hat \phi } \right)} \right]2\mathbb{Z} \nonumber\\
&  \left. + \left[ {\frac{{\lambda \left( {\lambda  + 1} \right) - \mu \left( {\mu  + 1} \right)}}{2}{Y_{\lambda \mu }}\left( {\hat \theta ,\hat \phi } \right) - \frac{{\lambda \left( {\lambda  + 1} \right) - \mu \left( {\mu  - 1} \right)}}{2}Y_{\lambda \mu }^*\left( {\hat \theta ,\hat \phi } \right)} \right]\mathbb{W}  \right\}  \nonumber\\
&  + B\sum\limits_{k\lambda \mu } {{{\left[ {{U_\lambda }\left( k \right)} \right]}^2}} {\mu ^2}{Y_{\lambda \mu }}\left( {\hat \theta ,\hat \phi } \right)Y_{\lambda \mu }^*\left( {\hat \theta ,\hat \phi } \right)2\mathbb{Z}  + 2B\sum\limits_{k\lambda \mu } {{{\left[ {{U_\lambda }\left( k \right)} \right]}^2}} \frac{{\lambda \left( {\lambda  + 1} \right) - \mu \left( {\mu  - 1} \right)}}{2}{Y_{\lambda \mu  - 1}}\left( {\hat \theta ,\hat \phi } \right)Y_{\lambda \mu  - 1}^*\left( {\hat \theta ,\hat \phi } \right)2\mathbb{Z} \nonumber\\
& + \sum\limits_{k\lambda \mu } {{\omega _k}} {\left[ {{U_\lambda }\left( k \right)} \right]^2}\left\{ {2{Y_{\lambda \mu }}\left( {\hat \theta ,\hat \phi } \right)Y_{\lambda \mu }^*\left( {\hat \theta ,\hat \phi } \right)\mathbb{Z}  + \left[ {{Y_{\lambda \mu }}\left( {\hat \theta ,\hat \phi } \right) + Y_{\lambda \mu }^*\left( {\hat \theta ,\hat \phi } \right)} \right]\mathbb{W} } \right\} \nonumber\\
& - \sum\limits_{k\lambda \mu } {{{\left[ {{U_\lambda }\left( k \right)} \right]}^2}\left[ {Y_{\lambda \mu }^*\left( {\hat \theta ,\hat \phi } \right){Y_{\lambda \mu }}\left( {\hat \theta ,\hat \phi } \right) + {Y_{\lambda \mu }}\left( {\hat \theta ,\hat \phi } \right)Y_{\lambda \mu }^*\left( {\hat \theta ,\hat \phi } \right)} \right]} \left| {LM} \right\rangle=0,\tag{B1}
\end{align}
\end{figure}
\begin{align}
&\frac{{\partial \left\langle 0_{\rm ph} \right|\langle LM|{{\tilde H}_0}\left| {LM} \right\rangle \left| 0_{\rm ph} \right\rangle }}{{\partial \mathbb{W}}}
= \left\langle 0_{\rm ph} \right|\left\langle {LM\left| {\frac{{\partial {{\tilde H}_0}}}{{\partial \mathbb{W}}}} \right|LM} \right\rangle \left| 0_{\rm ph} \right\rangle  \nonumber\\
=& \langle LM|B\sum\limits_{k\lambda \mu } {{{\left[ {{U_\lambda }\left( k \right)} \right]}^2}\mu \left[ {{Y_{\lambda \mu }}\left( {\hat \theta ,\hat \phi } \right) + Y_{\lambda \mu }^*\left( {\hat \theta ,\hat \phi } \right)} \right]} \mathbb{Z} \left( {{{\hat M}_0} - {{\hat J}_0}} \right)\nonumber\\
& - \frac{B}{2}\sum\limits_{k\lambda \mu } {{{\left[ {{U_\lambda }\left( k \right)} \right]}^2}\mu \left[ {{Y_{\lambda \mu }}\left( {\hat \theta ,\hat \phi } \right) + Y_{\lambda \mu }^*\left( {\hat \theta ,\hat \phi } \right)} \right]} \mathbb{Z}  \nonumber\\
& - \frac{B}{2}\sum\limits_{k\lambda \mu } {{{\left[ {{U_\lambda }\left( k \right)} \right]}^2}} {\mu ^2}\left[ {Y_{\lambda \mu }^*\left( {\hat \theta ,\hat \phi } \right) - {Y_{\lambda \mu }}\left( {\hat \theta ,\hat \phi } \right)} \right]\mathbb{Z} \nonumber\\
&- 2B\left\{ \frac{1}{2}\sum\limits_{k\lambda \mu } {{{\left[ {{U_\lambda }\left( k \right)} \right]}^2}} \left[ {\sqrt {\frac{{\lambda \left( {\lambda  + 1} \right) - \mu \left( {\mu  - 1} \right)}}{2}} {Y_{\lambda \mu  - 1}}\left( {\hat \theta ,\hat \phi } \right) + \sqrt {\frac{{\lambda \left( {\lambda  + 1} \right) - \mu \left( {\mu  + 1} \right)}}{2}} Y_{\lambda \mu  + 1}^*\left( {\hat \theta ,\hat \phi } \right)} \right]\mathbb{Z} \left( {{{\hat M}_{ + 1}} - {{\hat J}_{ + 1}}} \right)\right. \nonumber\\
& - \frac{1}{2}\sum\limits_{k\lambda \mu } {{{\left[ {{U_\lambda }\left( k \right)} \right]}^2}} \left[ {\sqrt {\frac{{\lambda \left( {\lambda  + 1} \right) - \mu \left( {\mu  + 1} \right)}}{2}} {Y_{\lambda \mu  + 1}}\left( {\hat \theta ,\hat \phi } \right) + \sqrt {\frac{{\lambda \left( {\lambda  + 1} \right) - \mu \left( {\mu  - 1} \right)}}{2}} Y_{\lambda \mu  - 1}^*\left( {\hat \theta ,\hat \phi } \right)} \right]\mathbb{Z} \left( {{{\hat M}_{ - 1}} - {{\hat J}_{ - 1}}} \right) \nonumber\\
& \left.+ \frac{1}{2}\sum\limits_{k\lambda \mu } {{{\left[ {{U_\lambda }\left( k \right)} \right]}^2}} \left[ {\frac{{\lambda \left( {\lambda  + 1} \right) - \mu \left( {\mu  + 1} \right)}}{2}{Y_{\lambda \mu }}\left( {\hat \theta ,\hat \phi } \right) - \frac{{\lambda \left( {\lambda  + 1} \right) - \mu \left( {\mu  - 1} \right)}}{2}Y_{\lambda \mu }^*\left( {\hat \theta ,\hat \phi } \right)} \right]\mathbb{Z} \right\} \nonumber\\
&  + \sum\limits_{k\lambda \mu } {{\omega _k}} {\left[ {{U_\lambda }\left( k \right)} \right]^2}\left\{ {\left[ {Y_{\lambda \mu }^*\left( {\hat \theta ,\hat \phi } \right) + {Y_{\lambda \mu }}\left( {\hat \theta ,\hat \phi } \right)} \right]\mathbb{Z}  + 2\mathbb{W} } \right\}- \sum\limits_{k\lambda \mu } {{{\left[ {{U_\lambda }\left( k \right)} \right]}^2}\left[ {Y_{\lambda \mu }^*\left( {\hat \theta ,\hat \phi } \right) + {Y_{\lambda \mu }}\left( {\hat \theta ,\hat \phi } \right)} \right]} \left| {LM} \right\rangle =0.\tag{B2}
\end{align}
By simplifying the Eqs. (B1) and (B2) further, we get the equations
\begin{align}
\left\{ {B\left[ {\left( {{\mu ^2} - \mu } \right){\mathfrak{T}_0} - \frac{{\lambda \left( {\lambda  + 1} \right) - \mu \left( {\mu  + 1} \right)}}{2}{\mathfrak{T}_{ + 1}} + 3\frac{{\lambda \left( {\lambda  + 1} \right) - \mu \left( {\mu  - 1} \right)}}{2}{\mathfrak{T}_{ - 1}}} \right] + {\omega _k}{\mathfrak{T}_0}} \right\}\mathbb{Z} + {\omega _k}\mathfrak{X}\mathbb{W} = {\mathfrak{T}_0},\tag{B3}
\end{align}
\begin{align}
\left( { - B\mu  + 2{\omega _k}} \right)\mathfrak{X}\mathbb{Z} + 2{\omega _k}\mathbb{W} = 2\mathfrak{X},\tag{B4}
\end{align}
where $\mathfrak{X} = \left\langle {LM} \right|{Y_{\lambda \mu }}\left| {LM} \right\rangle$, ${\mathfrak{T}_0} = \left\langle {LM} \right|{\left| {{Y_{\lambda \mu }}} \right|^2}\left| {LM} \right\rangle $, ${\mathfrak{T}_{ - 1}} = \left\langle {LM} \right|{\left| {{Y_{\lambda \mu  - 1}}} \right|^2}\left| {LM} \right\rangle $ and ${\mathfrak{T}_{ + 1}} = \left\langle {LM} \right|{\left| {{Y_{\lambda \mu  + 1}}} \right|^2}\left| {LM} \right\rangle$.
Solving the above equations, $\mathbb{Z}$ and $\mathbb{W}$ can be written as
\begin{align}
\mathbb{Z} = \frac{{2{\mathfrak{T}_0} - 2{\mathfrak{X}^2}}}{{\left[ {2B\left( {{\mu ^2} - \mu } \right) + 2{\omega _k}} \right]{\mathfrak{T}_0} - B\left[ {\lambda \left( {\lambda  + 1} \right) - \mu \left( {\mu  + 1} \right)} \right]{\mathfrak{T}_{ + 1}} + 3B\left[ {\lambda \left( {\lambda  + 1} \right) - \mu \left( {\mu  - 1} \right)} \right]{\mathfrak{T}_{ - 1}} + \left( {B\mu  - 2{\omega _k}} \right){\mathfrak{X}^2}}},\tag{B5}
\end{align}
\begin{align}
\mathbb{W} = \frac{{\mathfrak{X}\left\{ {2B{\mu ^2}{\mathfrak{T}_0} - B\left[ {\lambda \left( {\lambda  + 1} \right) - \mu \left( {\mu  + 1} \right)} \right]{\mathfrak{T}_{ + 1}} + 3B\left[ {\lambda \left( {\lambda  + 1} \right) - \mu \left( {\mu  - 1} \right)} \right]{\mathfrak{T}_{ - 1}}} \right\}}}{{{\omega _k}\left\{ {\left[ {2B\left( {{\mu ^2} - \mu } \right) + 2{\omega _k}} \right]{\mathfrak{T}_0} - B\left[ {\lambda \left( {\lambda  + 1} \right) - \mu \left( {\mu  + 1} \right)} \right]{\mathfrak{T}_{ + 1}} + 3B\left[ {\lambda \left( {\lambda  + 1} \right) - \mu \left( {\mu  - 1} \right)} \right]{\mathfrak{T}_{ - 1}} + \left( {B\mu  - 2{\omega _k}} \right){\mathfrak{X}^2}} \right\}}}.\tag{B6}
\end{align}
For the dominate channel of the phonon angular momentum $\lambda=2$ and the first-excited rotational state of angulon, namely, ${Y_{\lambda \mu }}\left( {\hat \theta ,\hat \phi } \right) = {Y_{20}}\left( {\hat \theta ,\hat \phi } \right)$ and $\left| {LM} \right\rangle \left. {\left| {{0_{\rm {ph}}}} \right.} \right\rangle = \left| {10} \right\rangle \left. {\left| {{0_{\rm {ph}}}} \right.} \right\rangle $, Eqs. (B5) and (B6) covert into
\begin{align}
\left( {6B\gamma  + {\omega _k}\alpha } \right)\mathbb{Z}  + {\omega _k}\chi \mathbb{W} = \alpha,\tag{B7}
\end{align}
\begin{align}
{\omega _k}\chi \mathbb{Z}  + {\omega _k}\mathbb{W}  = \chi.\tag{B8}
\end{align}
From them, the expressions for $\mathbb{Z}$ and $\mathbb{W}$ can be obtained as given in the main text.
\end{appendix}

\end{widetext}

\end{document}